\documentclass[twocolumn,secnumarabic,amssymb, nobibnotes, aps, prb, superscriptaddress]{revtex4-2}
\usepackage{lipsum}
\usepackage{titlesec}
\titlespacing\section{0pt}{12pt plus 4pt minus 4pt}{1pt plus 20pt minus 2pt}
\usepackage{xcolor}
\usepackage{tabularx}
\usepackage{algorithm}
\usepackage{amsmath}
\usepackage{comment}
\usepackage{physics}
\usepackage{afterpage}
\usepackage{placeins}
\usepackage{graphicx}
\usepackage{float}
\usepackage{booktabs}
\usepackage{multirow}
\usepackage{array}
\usepackage{setspace}
\graphicspath{{Figs/}}
\usepackage{siunitx}
\usepackage{csquotes}
\usepackage{hhline}
\usepackage{balance}
\usepackage{xfrac}
\usepackage{float,graphicx}
\usepackage{mathtools}
\usepackage{listings}
\usepackage{amssymb}
\usepackage{titlesec}
\usepackage{amsfonts}
\usepackage[version=4]{mhchem}

\usepackage{epstopdf}
\catcode`@11
\def\seceqaa{\@addtoreset{equation}{section}
\def\theequation{A\arabic{equation}}}
\def\seceqbb{\@addtoreset{equation}{section}
\def\theequation{B\arabic{equation}}}
\def\seceqcc{\@addtoreset{equation}{section}
\def\theequation{C\arabic{equation}}}
\def\seceqdd{\@addtoreset{equation}{section}
\def\theequation{D\arabic{equation}}}
\def\seceqee{\@addtoreset{equation}{section}
\def\theequation{E\arabic{equation}}}
\def\seceqff{\@addtoreset{equation}{section}
\def\theequation{F\arabic{equation}}}
\def\seceqgg{\@addtoreset{equation}{section}
\def\theequation{G\arabic{equation}}}
\def\seceqhh{\@addtoreset{equation}{section}
\def\theequation{H\arabic{equation}}}
\catcode`@11

\begin{document}
\title{Emergence of Spin Ice freezing in Dy$_2$Ti$_{1.8}$Mn$_{0.2}$O$_7$} 

\author{Rajnikant Upadhyay}
\affiliation{School of Materials Science and Technology, Indian Institute of Technology (Banaras Hindu University), Varanasi 221005, India}

\author{Manjari Shukla}
\affiliation{School of Materials Science and Technology, Indian Institute of Technology (Banaras Hindu University), Varanasi 221005, India}

\author{Rachana Sain}
\affiliation{School of Materials Science and Technology, Indian Institute of Technology (Banaras Hindu University), Varanasi 221005, India}

\author{Martin Tolkiehn}
\affiliation{Deutsches Elektronen-Synchrotron (DESY), Notkestraße 85, 22607, Hamburg, Germany}

\author{Chandan Upadhyay}
\affiliation{School of Materials Science and Technology, Indian Institute of Technology (Banaras Hindu University), Varanasi 221005, India}

\begin{abstract}
We herein present the spin freezing dynamics of stuffed polycrystalline compound  Dy$_2$Ti$_{1.8}$Mn$_{0.2}$O$_7$. In Dy$_2$Ti$_2$O$_7$, spin freezes with ice like spin relaxations at a temperature around 3 K (T$_i$) along with another spin freezing at a temperature around 0.7 K (T\textless T$_i$). These relaxations can be observed prominently with an application of varying DC magnetic field bias and applied AC-field. We show here that with fractional inclusion of Mn at Ti site in Dy$_2$Ti$_2$O$_7$, there is a significant shift in these temperatures. In Dy$_2$Ti$_{1.8}$Mn$_{0.2}$O$_7$ the T$_i$ shifts to a higher temperature around 5 K and freezing belonging to T\textless T$_i$ shifts to 2.5 K without any application of external DC Bias and/or AC-field. The inclusion of Mn at Ti site also enhances the ferromagnetic interaction for Dy$_2$Ti$_{1.8}$Mn$_{0.2}$O$_7$ as compared to Dy$_2$Ti$_2$O$_7$. Arrhenius fit of freezing temperature with frequency for Dy$_2$Ti$_{1.8}$Mn$_{0.2}$O$_7$  shows that these spin relaxations at T$_i$ and T\textless T$_i$  are thermally induced. Low-temperature structural change in lattice parameters and crystal field phonon coupling has been studied using synchrotron x-ray diffraction. Debye-Gruineisen analysis of temperature-dependent lattice volume shows the emergence of crystal field phonon coupling at a much higher temperature (70 K) in Dy$_2$Ti$_{1.8}$Mn$_{0.2}$O$_7$ in contrast to 40 K in Dy$_2$Ti$_2$O$_7$. These findings make Dy$_2$Ti$_{1.8}$Mn$_{0.2}$O$_7$ a suitable system to explore the application of spin ice phenomenon at a workable temperature.

\end{abstract}
\maketitle
\section{INTRODUCTION}
 Geometric frustration in a magnetic system occurs when the spatial arrangement of the magnetic moments (\enquote{spins}) combined with the magnetic interactions prevents the formation of a \enquote{simple} ordered  ground state. Rare earth pyrochlores A$_2$B$_2$O$_7$, where A is a trivalent ion and B is tetravalent ion,belongs to class of frustrated materials in which geometry of the magnetic sub-lattice leads to frustration \cite{subramanian1993rare}. It displays a wide variety of unconventional and exciting magnetic ground states from spin liquid to spin glass through spin ice behaviour \cite{gardner2010magnetic,greedan2006frustrated,balents2010spin}.
  For the spin ice compounds the near neighbour interactions between the Ising spins are dipolar ferromagnetic(FM) and weaker antiferromagnetic (AFM) exchange. Overall, this configuration results in effective frustrated ferromagnetic interaction in order to minimize the dipolar and ferromagnetic exchange interaction to attain the ground state, spins on each tetrahedron adopt the two-in-two-out configuration, which is analogous to the two-short-two-long proton bond configuration in water ice \cite{harris1997geometrical}.
These interaction results in six-fold degenerate ground state, which gives residual measurable zero-point entropy S$_o$ = R/2ln(3/2), where R is the molar gas constant \cite{bramwell2001spin}.A brief history of spin ice has been summarised by Bramwell and Harris \cite{bramwell2020history}
 
 Spin ice materials Dy$_2$Ti$_2$O$_7$, Ho$_2$Ti$_2$O$_7$, Dy$_2$Sn$_2$O$_7$ and Ho$_2$Sn$_2$O$_7$, in which lattice geometry leads to frustration due to competing ferromagnetic and dipolar interactions, have shown a rich physics originating from its inherent geometric frustration and resultant highly degenerate ground state \cite{ramirez1999zero,siddharthan1999ising,bramwell2001spin}.The structure is represented by the space group Fd$\Bar{3}$m featuring two sub-lattices that interpenetrate each other and consist of networks of corner-sharing tetrahedra. The magnetic rare-earth ion resides on the lattice of corner-sharing tetrahedra, where spin are constrained by crystal field to point either directly toward or directly away from the centre of the tetrahedra.
Specifically, for the pyrochlores with Dy and Ho at the A site, the rare-earth ion is subject to a strong crystal-electric-field splitting of the ground state and the first excited state of the single ion by an energy of the order of 200 K \cite{matsuhira2001novel,rosenkranz2000crystal}. In Dy$_2$Ti$_2$O$_7$, three relaxation phenomena have been observed, associated with single-spin relaxation at higher temperatures and spin with ice like correlations at lowest temperatures, but with quantum spin relaxation observed at intermediate temperatures. \cite{snyder2001spin,snyder2004low,snyder2003quantum,snyder2004quantum}.

To examine the development of the magnetic ground state of stuffed spin ice. In the present study, we have used dc and ac magnetic susceptibility to investigate the nature of magnetic interaction and spin relaxation in thermal and quantum spin relaxation regimes by inclusion of small fraction of Mn at the Ti site, in the spin ice compound Dy$_2$Ti$_2$O$_7$ within the domain of phase space. It has been shown that spin ice freezing temperature shifts to higher temperature with the inclusion of Mn. This suggests that a relatively low level of inclusion of Mn at Ti site changing the number and type of relaxation processes available to the spin and freezing temperature in such system, making it interesting to study how low-temperature magnetic interaction evolves with Mn substitution at Ti site in Dy$_2$Ti$_2$O$_7$.
\section{EXPERIMENTAL}
Polycrystalline Dy$_2$Ti$_2$O$_7$ and Dy$_2$Ti$_{1.8}$Mn$_{0.2}$O$_7$ samples were prepared using the standard solid-state thermochemical reaction method \cite{shukla2020robust}. The starting materials for synthesis was Dy$_2$O$_3$, TiO$_2$ and MnO$_2$ sample was heated in ambient conditions at 1400 °C with multiple heating and grinding until the reaction was complete. High-resolution X-ray diffraction demonstrated the samples to be in single-phase pyrochlore. We measured dc susceptibility, as well as the real and imaginary part of ac susceptibility, $\chi\prime (T)$ and $\chi\prime\prime(T)$ applying excitation field H$_{ac}$ = 2.5 Oe at different dc bias field using Quantum design MPMS superconducting quantum interference device magnetometer. Synchrotron x-ray diffraction (SXRD) has been carried out to investigate the structural changes at low temperatures using P24 beamline at Petra III at DESY, Hamburg, Germany. The diffraction pattern was obtained using a 25 KeV x-ray radiation source in the temperature range of (15-140) K with a Q value ranging from 0.2 to 7.5. Rietveld refinement for the obtained SXRD was performed using FULLPROF SUIT \cite{rodriguez2001introduction}.
\section{RESULTS AND DISCUSSION}
The static magnetization was measured using zero field cooling and field cooling protocol in the temperature range from 2 K to 200 K at a rate of 2 K. 
\begin{figure}[htbp]
    \centering
    \includegraphics[width=0.50\textwidth]{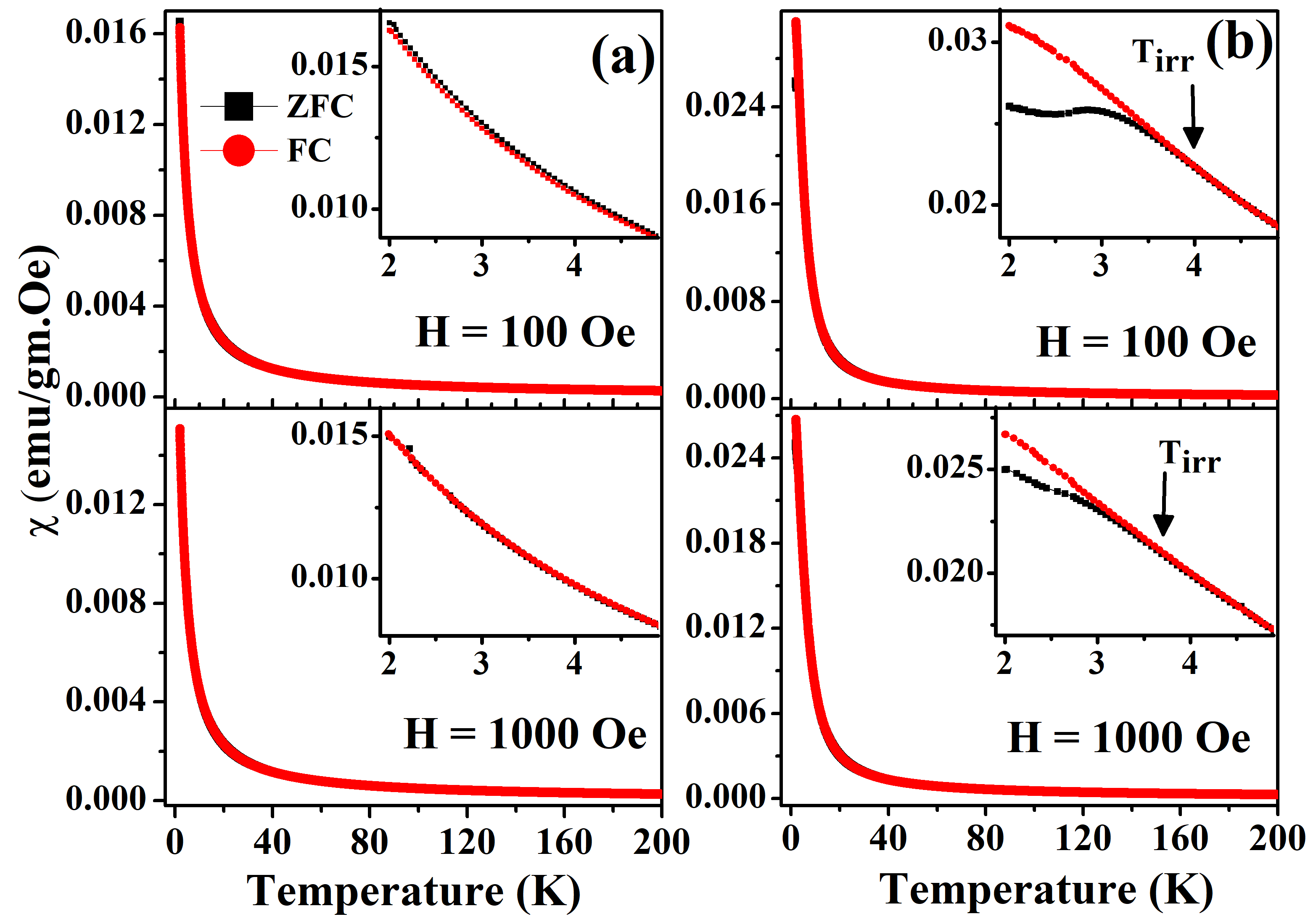}
    \caption{ Temperature dependence of magnetization measured in ZFC and FC protocol for (a) Dy$_2$Ti$_2$O$_7$ and (b)  Dy$_2$Ti$_{1.8}$Mn$_{0.2}$O$_7$  at an applied field of 100 Oe and 1000 Oe.}
    \label{Fig1}
\end{figure}
FIG.\ref{Fig1}(a) shows zero field cooled (ZFC) and field Cooled (FC) magnetization data for Dy$_2$Ti$_2$O$_7$. There is no difference in ZFC and FC curve in the measured temperature range, suggesting no magnetic ordering or absence of spin-glass like transition down up to 2 K as reported earlier \cite{liu2014magnetic}. However, the dc magnetic susceptibility shows clear bifurcation in ZFC and FC for Dy$_2$Ti$_{1.8}$Mn$_{0.2}$O$_7$ at a lower temperature around 3.8 K (FIG.\ref{Fig1}(b). The irreversibility in ZFC and FC signifies enhanced ferromagnetism which is expected because of increasing disorder \cite{liu2015enhanced}.
\begin{figure}[htbp]
    \centering
    \includegraphics[width=0.50\textwidth]{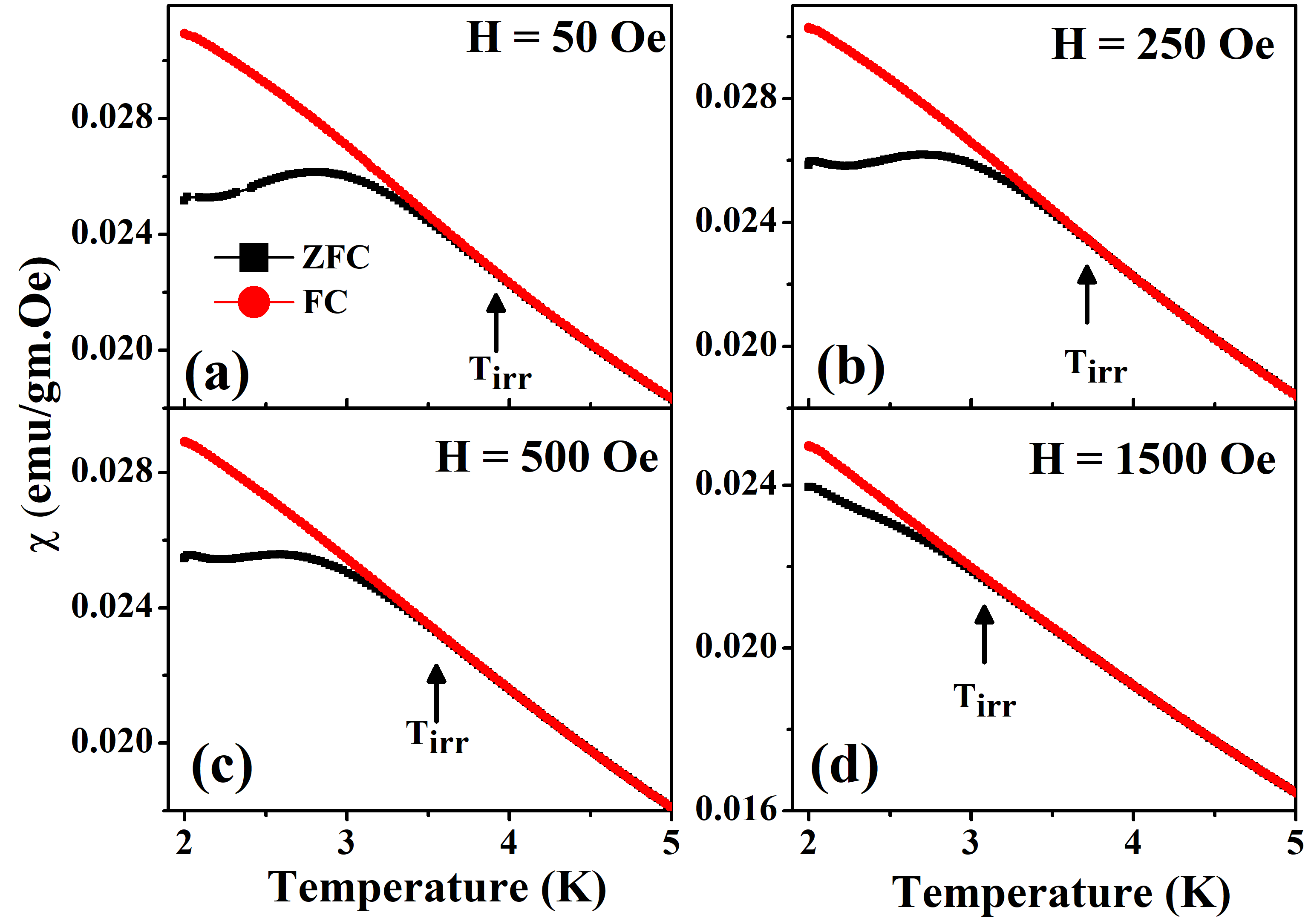}
    \caption{ Temperature dependence of magnetization measured in ZFC and FC protocol for Dy$_2$Ti$_{1.8}$Mn$_{0.2}$O$_7$ at applied field of (a) 50 Oe (b) 250 Oe (c) 500 Oe (d) 1500 Oe.}
    \label{Fig2}
\end{figure}
We have also seen the effects of applied magnetic field on freezing temperature i.e bifurcation point. FIG.\ref{Fig2} shows that bifurcation temperature (T$_{irr}$) decreases with an increase in the magnetic field from 50 Oe to 1500 Oe; as expected for materials lacking long-range magnetic ordering \cite{liu2015enhanced}. This behaviour is similar to typical spin-glass system i.e. monotonic decrease of ZFC curve below T$_{irr}$ and also with the increase in the applied field, the percentage difference between two ZFC and FC data set decreases. With further increase in the field above 1500 Oe no difference in ZFC and FC curve is observed \cite{snyder2004low}.
\begin{figure}[htbp]
    \centering
    \includegraphics[width=0.5\textwidth]{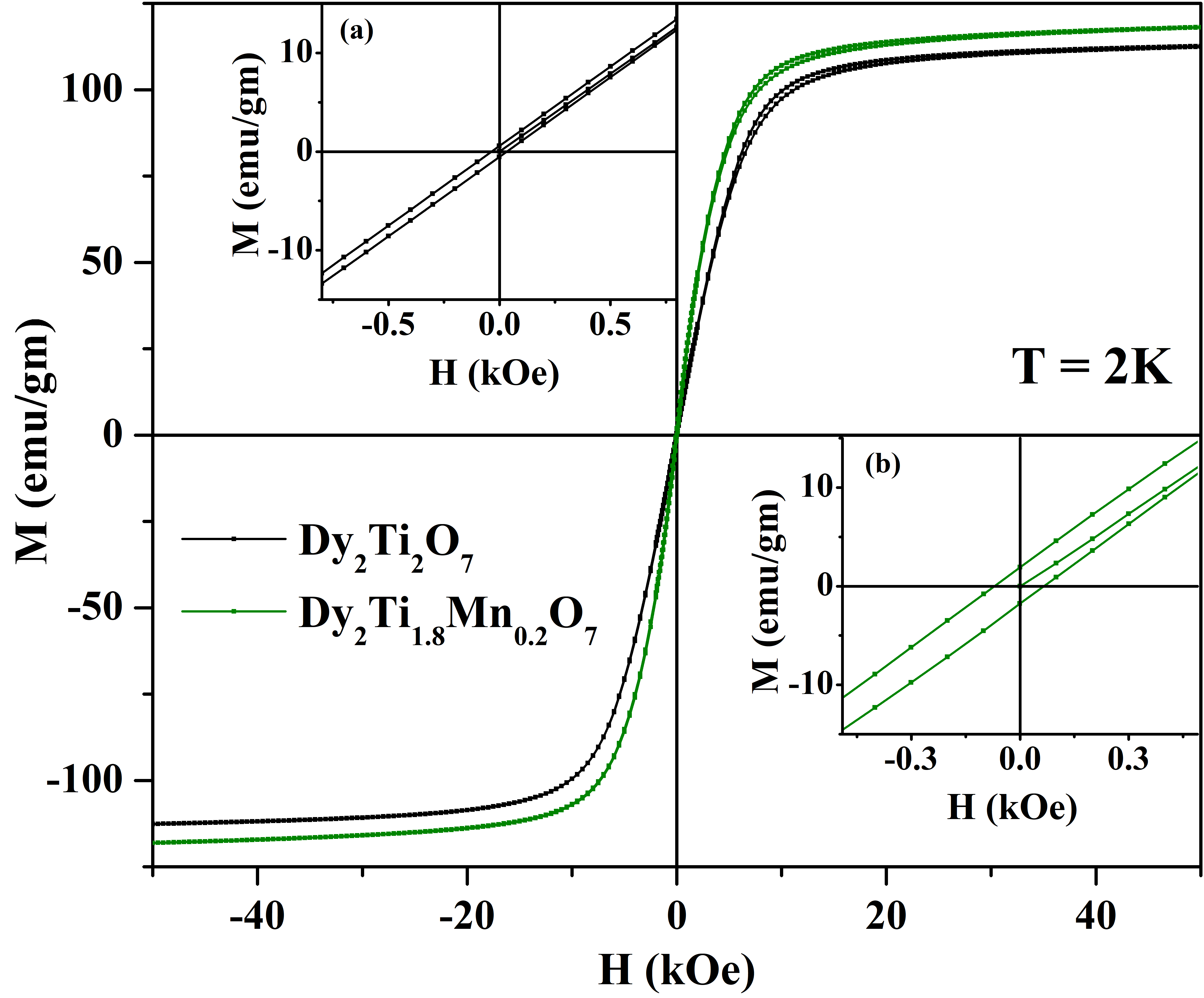}
    \caption{ Magnetization as a function of Field (MH) curve for Dy$_2$Ti$_2$O$_7$ and Dy$_2$Ti$_{1.8}$Mn$_{0.2}$O$_7$ at 2K for H (magnetic field) ranging from -50 to 50 kOe.}
    \label{Fig3}
\end{figure}

FIG.\ref{Fig3} show the field dependence of magnetization (M-H) curve measured at 2K; the field was swept upto 5 kOe, down to -5 kOe, and back upto 5 kOe to complete the hysteresis loop. M$_{max}$ value calculated from the M-H plot is 42.72 µ$_B$/unit cell  in the case of Dy$_2$Ti$_2$O$_7$, which is consistent with previously reported value. The M$_{max}$ value increases to 44.776  µ$_B$/unit cell for Dy$_2$Ti$_{1.8}$Mn$_{0.2}$O$_7$ \cite{anand2015investigations,fukazawa2002magnetic}. The obtained value is half of the actual magnetic moment of due to the angular averaging in case of the powdered sample. The saturation value of magnetization shows a slight increase in M$_{max}$ value for Dy$_2$Ti$_{1.8}$Mn$_{0.2}$O$_7$, indicating that inclusion does not measurably alter the system's anisotropy \cite{snyder2004quantum}. The magnetization increases upto 20 kOe and then remains saturated upto an applied field of 50 kOe.
\begin{figure}[t]
    \centering
    \includegraphics[width=0.5\textwidth]{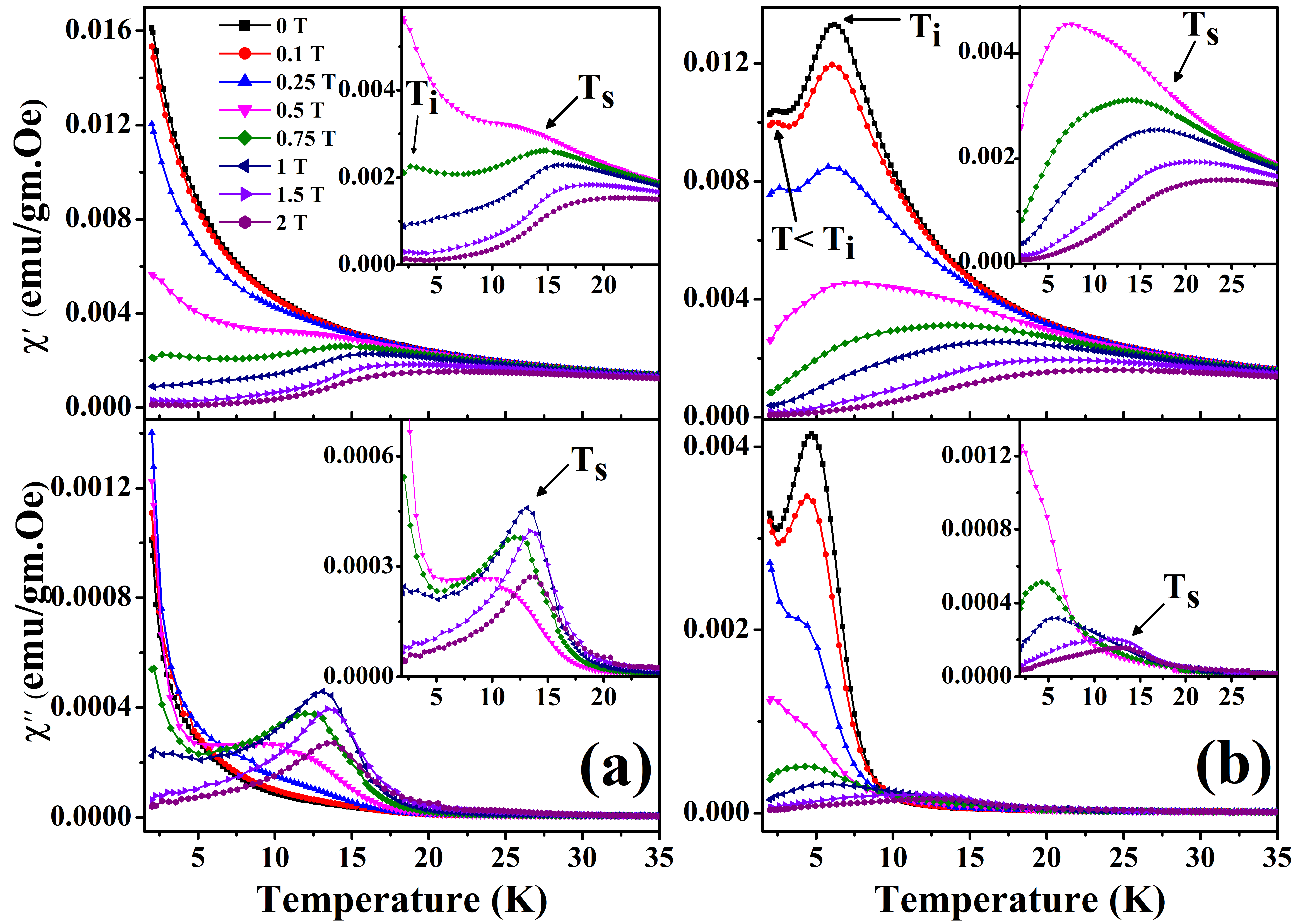}
    \caption{ The real (upper panel) and imaginary part (lower panel) of ac susceptibility at several applied dc magnetic fields and at frequency of 300 Hz for (a) Dy$_2$Ti$_2$O$_7$ (b) Dy$_2$Ti$_{1.8}$Mn$_{0.2}$O$_7$.}
    \label{Fig4}
\end{figure}

The evidence of bifurcation in the MT plot for Dy$_2$Ti$_{1.8}$Mn$_{0.2}$O$_7$ at measured temperature made us probe into the system's spin dynamics. Since there is a bifurcation in the dc susceptibility plot, we expect freezing above this temperature in ac susceptibility data.
FIG.\ref{Fig4} shows real and imaginary part of field-dependence ac susceptibility measurement at a frequency of 300 Hz and DC bias magnetic field ranging from 0-2 T.
\begin{figure}[b]
    \centering
    \includegraphics[width=0.5\textwidth]{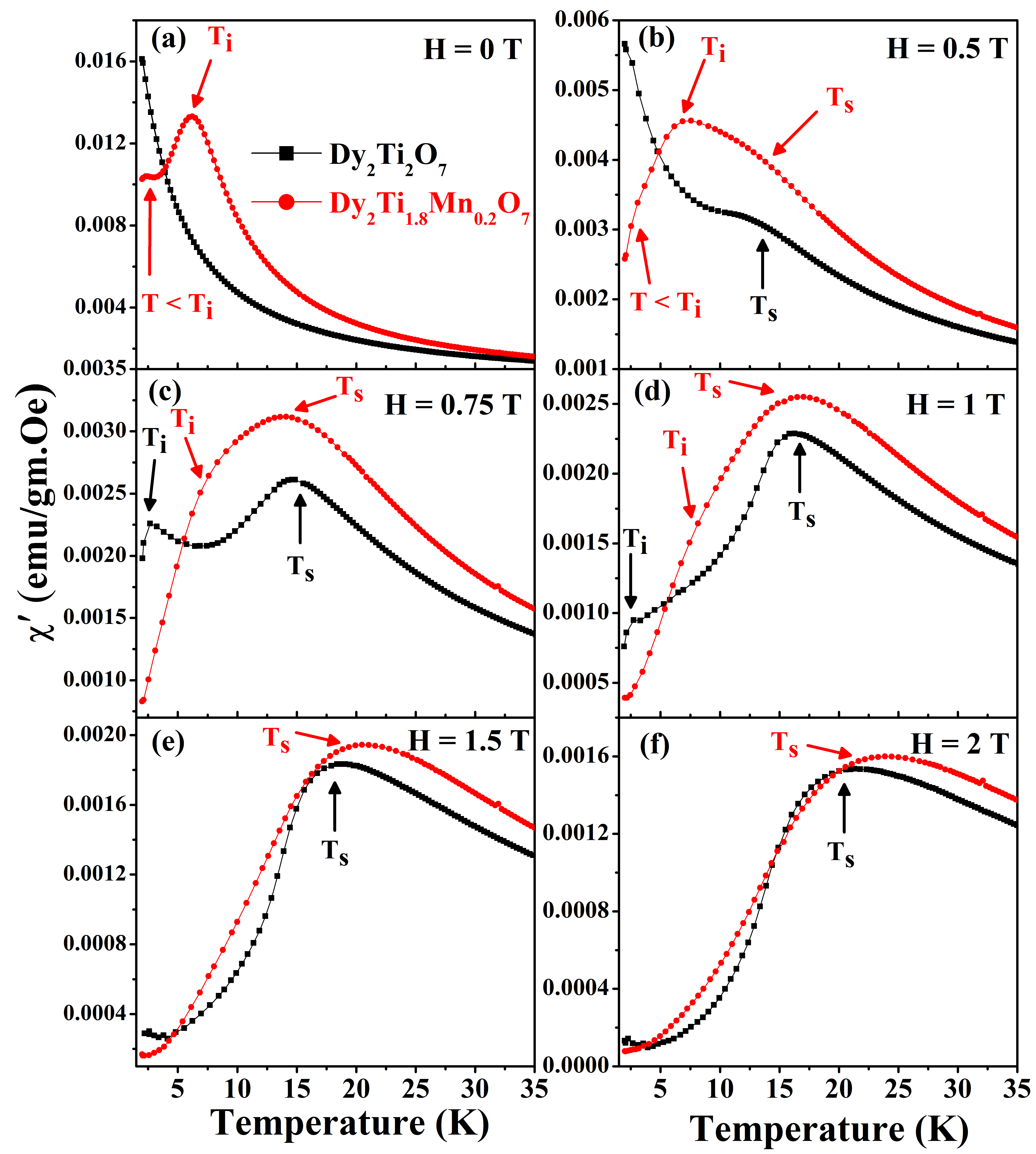}
    \caption{ Temperature dependence of the real part of ac susceptibility measured at the 300 Hz frequency for Dy$_2$Ti$_2$O$_7$ and Dy$_2$Ti$_{1.8}$Mn$_{0.2}$O$_7$ at different DC bias fields (a) 0 T (b) 0.5 T (c) 0.75 T(d) 1 T (e) 1.5 T (f) 2 T}
    \label{Fig5}
\end{figure}
The real part of ac susceptibility is virtually identical to dc susceptibility in case of Dy$_2$Ti$_2$O$_7$, in the absence of an applied dc field, FIG.\ref{Fig4}(a). With increasing the field after 0.25 T we see the emergence of two relaxations, one corresponding to single ion freezing temperature (T$_s$)  at around 15 K \cite{snyder2004quantum} and a spin ice state that emerges around 3 K (T$_i$) \cite{ke2007nonmonotonic} appears in real part of ac susceptibility  $\chi\prime (T)$, we also observe peak in $\chi\prime\prime(T)$ corresponding to maximum in $\chi\prime (T)$ as expected from Kramer-Kronig relation \cite{snyder2004low,xing2010emergent}. The freezing at higher temperature become more pronounced with increase in the field which is the characteristics for spin ice, whereas for spin glasses and superparamagnets magnetic field suppresses the freezing temperature \cite{gardner1999glassy,mydosh1993spin}. The single-ion freezing (T$_s$) shifts towards higher temperature with increase in magnetic field. Such behavior has been observed by Shi et al. in ac susceptibility measurements performed on single crystal of Dy$_2$Ti$_2$O$_7$ \cite{shi2007dynamical}.
\begin{table*}[htp]
\caption{\label{Table:1}The parameters $V (T=0)$, $\theta_{D}$ and 9$\gamma$Nk$_{B}$/B  for Dy$_2$Ti$_2$O$_7$ and Dy$_2$Ti$_{1.8}$Mn$_{0.2}$O$_7$ obtained by fitting Debye Function.}
\begin{ruledtabular}
\begin{tabular}{ccccc}
    \textbf{Compound}& V (T = 0K) (\AA$^3$)& Debye temperature, $\theta_{D} (K)$ & 9$\gamma$Nk$_B$/B (\AA$^3$ /K) \\ \hline
     Dy$_2$Ti$_2$O$_7$ &1034.7  &214.1 &0.065 \\
     Dy$_2$Ti$_{1.8}$Mn$_{0.2}$O$_7$ &1032.3 &290.6 &0.066 \\
\end{tabular}
\end{ruledtabular}
\end{table*}
Whereas in case of Dy$_2$Ti$_{1.8}$Mn$_{0.2}$O$_7$, FIG.\ref{Fig4}(b), there appears two relaxation at a lower temperature near spin ice state even in the absence of dc field first one around 4 K (T$_i$) and another one below T$_i$ at around 2.5 K (T\textless T$_i$). The one which is around 2.5 K (T\textless T$_i$) can be correlated to the low temperature freezing which comes at around 0.7 K in Dy$_2$Ti$_2$O$_7$ which is studied in detail by Snyder et al. \cite{snyder2004low}. It Suggests that the inclusion of Mn shifts the spin ice freezing state to higher temperature.Further with increase in the field the single ion freezing (T$_s$) appears and the one at lowest temperature (T\textless T$_i$) disappears. Real part  of ac susceptibility $\chi\prime (T)$ shows suppression in higher temperature freezing (T$_s$) in case of Dy$_2$Ti$_{1.8}$Mn$_{0.2}$O$_7$ when we increase the field above 0.5 T.To analyze the role of Mn,ac susceptibility for both the compound are plotted in single figure at various applied field, FIG.\ref{Fig5}. FIG.\ref{Fig5}(a) shows spin ice freezing in Dy$_2$Ti$_{1.8}$Mn$_{0.2}$O$_7$ at zero dc bias which is not observed in Dy$_2$Ti$_2$O$_7$ .As shown in  FIG.\ref{Fig5}(b) T$_s$ is suppressed at an applied dc field of 0.5 T. As we increase the applied field to 0.75 T the freezing at lower temperature T$_i$ merges with T$_s$ and gives a single hump like feature FIG.\ref{Fig5}(c). Further with increase of field to 2 T both Dy$_2$Ti$_2$O$_7$  and Dy$_2$Ti$_{1.8}$Mn$_{0.2}$O$_7$ shows similar behaviour FIG.\ref{Fig5}(f). This shows that substitution of Mn alters the spin dynamics of the system at lower temperature regime i.e near spin ice state and higher temperature regime, near single ion freezing temperature in different way. 
\begin{figure}[htbp]
    \centering
    \includegraphics[width=0.5\textwidth,]{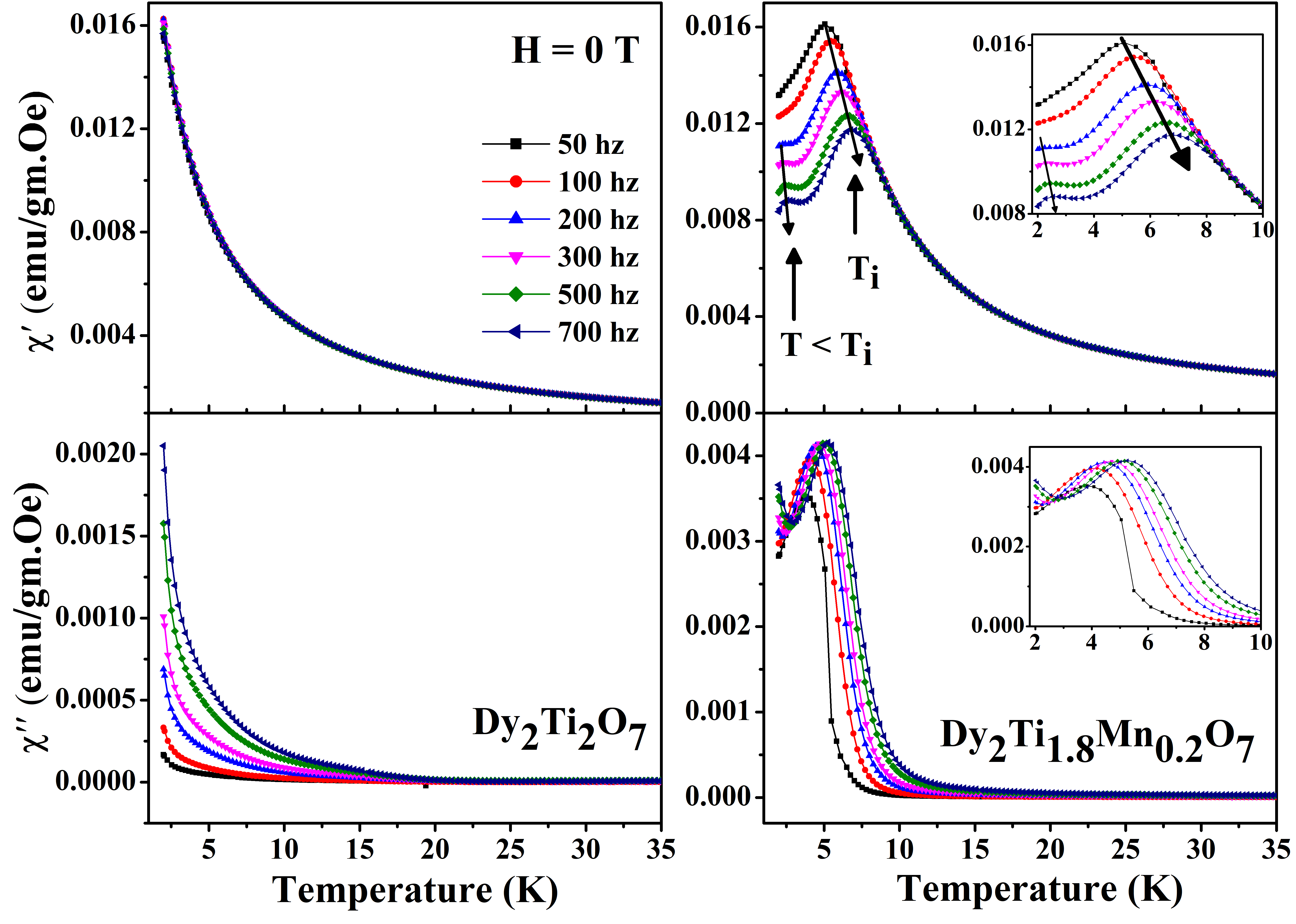}
    \caption{Frequency dependence for Dy$_2$Ti$_2$O$_7$ and Dy$_2$Ti$_{1.8}$Mn$_{0.2}$O$_7$ at an applied field of 0 T}
    \label{Fig6}
\end{figure}
\begin{figure}[htbp]
    \centering
    \includegraphics[width=0.5\textwidth,]{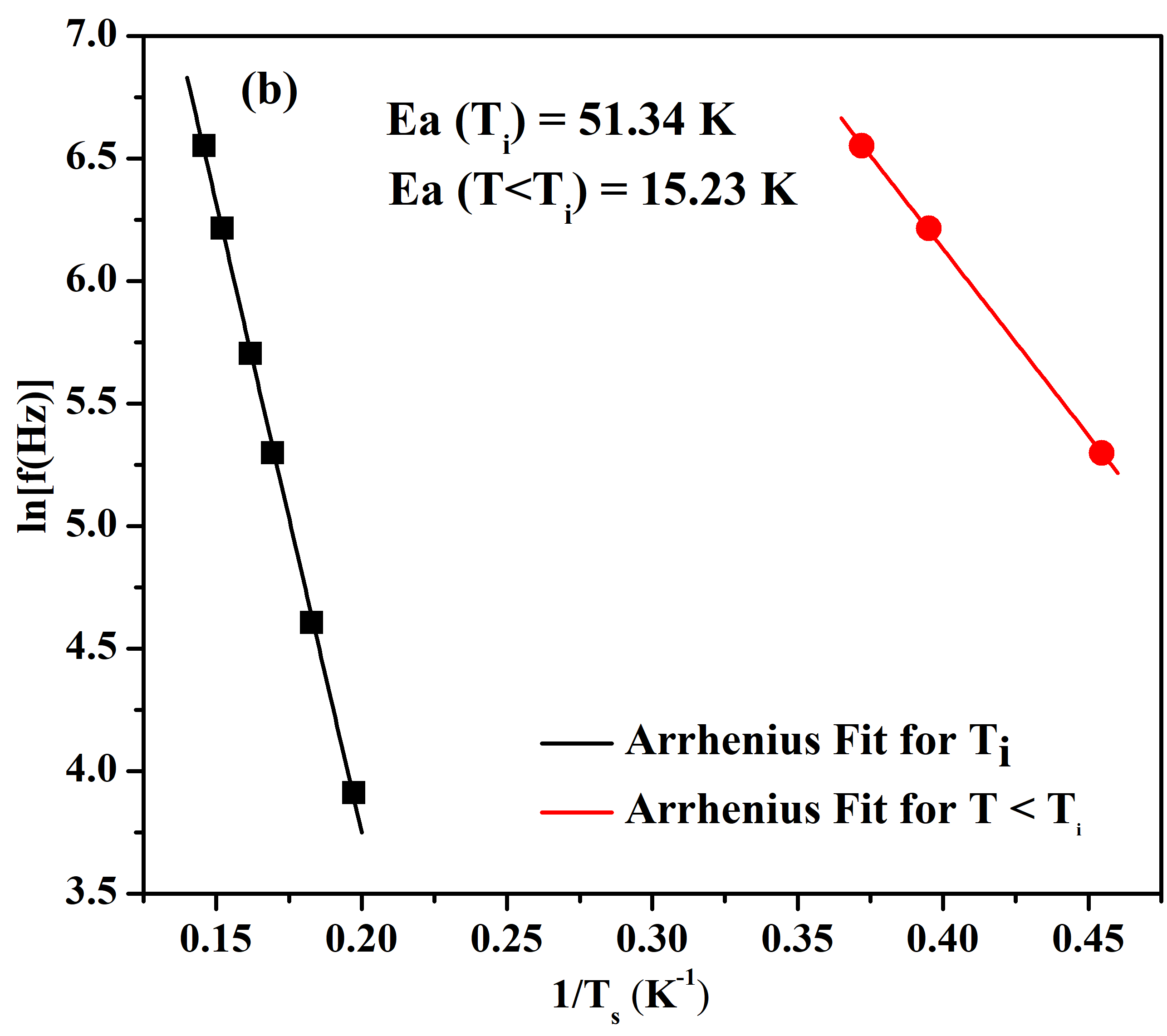}
    \caption{Arrhenius law fit of freezing temperatures (T$_i$ and T\textless T$_i$) dependence on frequency for Dy$_2$Ti$_{1.8}$Mn$_{0.2}$O$_7$}
    \label{Fig7}
\end{figure}

To examine the nature of freezing phenomena, ac susceptibility data is collected in the vicinity of low temperature at varying frequencies ranging from 50 Hz to 700 Hz at various applied dc fields. FIG.\ref{Fig6} shows the temperature dependent susceptibility as a function of frequency for both the compounds in absence of applied dc field. The real part of ac susceptibility measurement $\chi\prime (T)$ shows canonical paramagnetic type behaviour for Dy$_2$Ti$_2$O$_7$ for all applied frequencies (f = 50, 100, 200, 300, 500, 700 Hz) corresponding to which there is sharp increase in the imaginary part $\chi\prime\prime(T)$ in the lower panel as well.
Whereas for  Dy$_2$Ti$_{1.8}$Mn$_{0.2}$O$_7$, we see that both lower temperature spin ice freezing which is around 4 K (T$_i$) and 2.5 K (T\textless T$_i$) shows strong frequency dependence. To investigate the nature of T$_i$ and (T \textless T$_i$) as well as the spin dynamics in Dy$_2$Ti$_{1.8}$Mn$_{0.2}$O$_7$, we examine the frequency dependence of T$_i$ and (T\textless T$_i$) by fitting the data to an Arrhenius law given below
\begin{equation}
f = f_o e^\frac{E_{A}}{K_{B}T}
\end{equation}
                                                            
Where $E_{A}$ is the activation energy for spin fluctuations, $K_{B}$ is Boltzmann constant and $f_{0}$ is a measure of the microscopic limiting frequency in the system \cite{snyder2004quantum}. FIG.\ref{Fig7} shows the Arrhenius fit, the activation energy comes out to be 51.34 K for spin relaxation at T$_i$ and 15.23 K for the spin relaxation at T\textless T$_i$ in case of Dy$_2$Ti$_{1.8}$Mn$_{0.2}$O$_7$ , it seems that activation energy  has been increased for spin relaxation at T$_i$  which is expected as lattice constant has decreased \cite{snyder2004quantum}. The Arrhenius law behaviour indicates that spin relaxation are thermally driven \cite{snyder2003quantum}.
The values of the freezing temperature are obtained from the minimum in the slope of $\chi\prime (T)$ with Mn substitution T$_i$ and (T\textless T$_i$) shifts towards the higher temperature side with increase in frequency this peak shift towards higher temperature.
\begin{figure*}[htbp]
    \centering
    \includegraphics[width=0.95\textwidth,height=10cm]{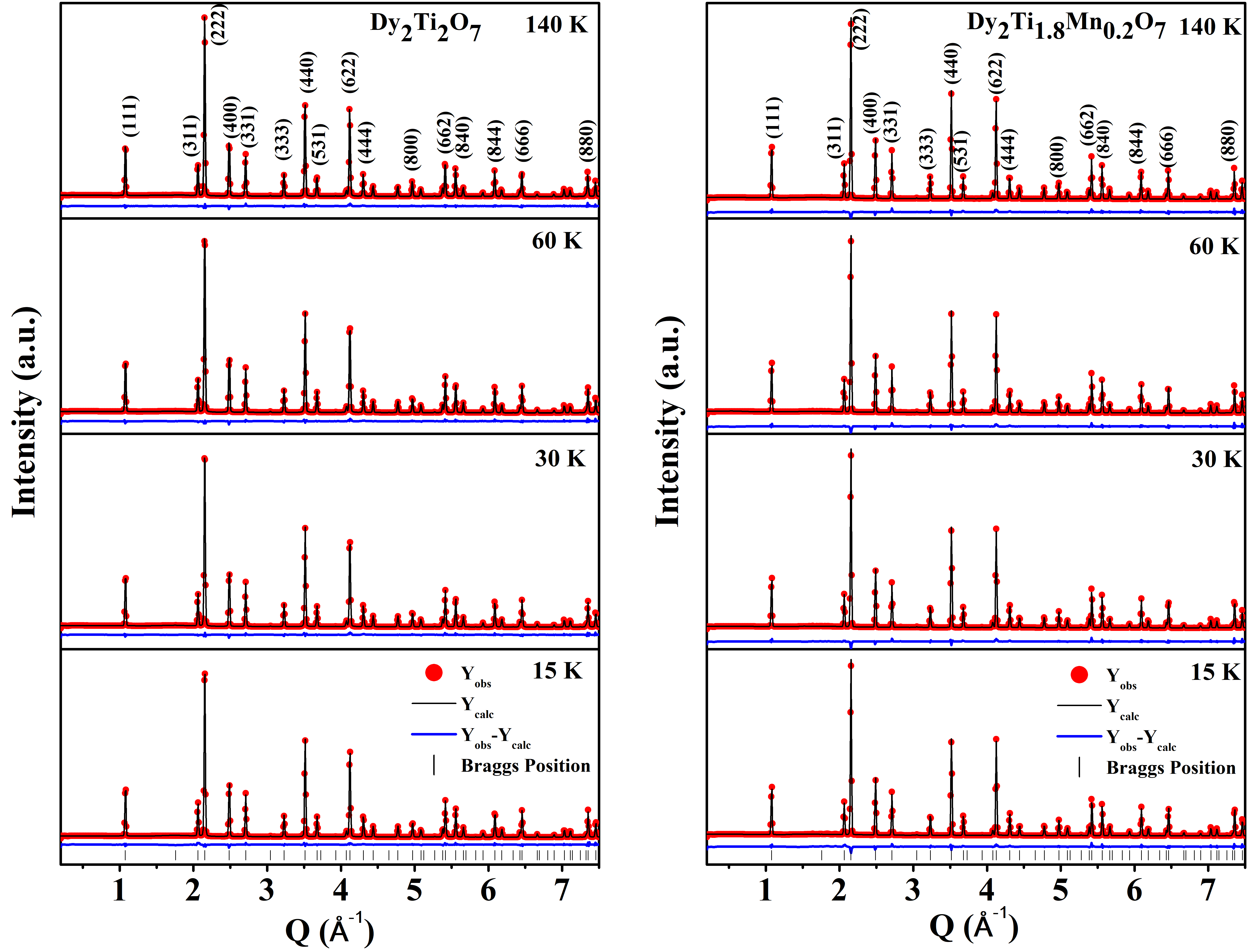}
    \caption{Rietveld refinement of low temperature synchrotron x-ray diffraction pattern
of Dy$_2$Ti$_2$O$_7$ and Dy$_2$Ti$_{1.8}$Mn$_{0.2}$O$_7$ at T = 140 K, T = 60 K, T = 30\,K and T = 15\,K.}
    \label{Fig8}
\end{figure*}
\begin{figure*}[htbp]
    \centering
    \includegraphics[width=0.95\textwidth]{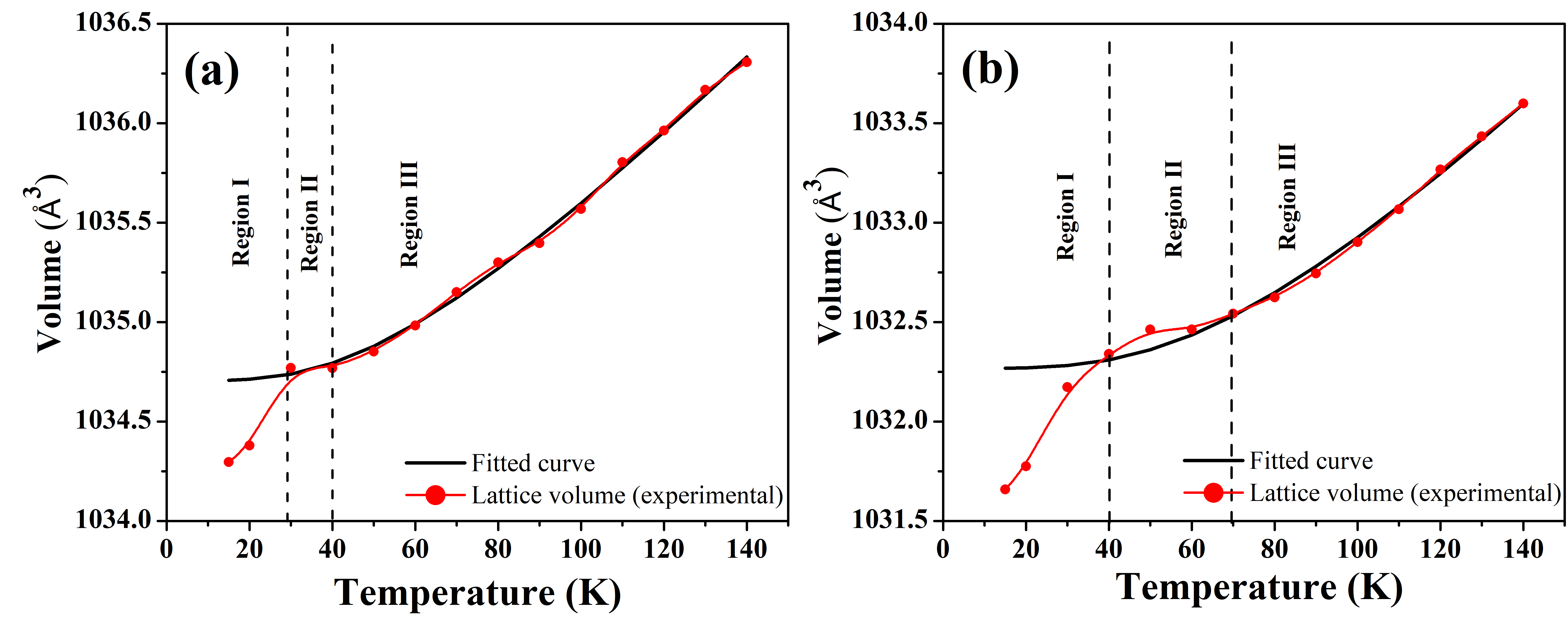}
    \caption{Temperature variations of unit cell parameters solid line (black) represent the fitted curve contribution of phonons by Debye Gruineisen, and solid circle (red) represents experimental lattice volume (a) Dy$_2$Ti$_2$O$_7$ and (b)Dy$_2$Ti$_{1.8}$Mn$_{0.2}$O$_7$.}
    \label{Fig9}
\end{figure*}

SXRD was performed from 140 K to 15 K at an interval of 10 K, to establish the Crystal field phonon coupling. FIG.\ref{Fig8} shows the some selected low-temperature SXRD pattern for T = 140 K, 60 K, 30 K, 15 K and its rietveld refienment.
Temperature-dependent variation of lattice volume has been plotted. Debye Gruineisen equation has been fitted to extract the crystal field-induced change in the lattice volume.
The equation is as follows:
\begin{equation*}
V \cong V\left (T = 0\right) + \int_0^T \frac{\gamma C_{\nu}}{B} dT
\end{equation*}

\begin{equation}
V \cong V (T = 0) + \frac{9 \gamma N k_{B}}{B}  T \left (\frac{T}{\theta_{D}}
\right)^3 \int_0^\frac{\theta_D}{T} \frac{e^x}{e^x-1} dx  
\end{equation}

V (T=0), $\theta_{D}$ and 9$\gamma$Nk$_{B}$/B are fitting parameters; V (T = 0) represents the lattice volume at 0 K, $\theta_{D}$ is the Debye temperature, $\gamma$ is Grüneisen parameter and B is the bulk modulus \cite{yakub2017modified,kiyama1996invar}.
Debye model accounts for contribution to thermal expansion due to anharmonic parts of lattice vibration. As shown in FIG.\ref{Fig9}(a) lattice volume decreases linearly till 40 K, below which it shows apparent deviation from Debye Gruineisen fit, showing the dominance of crystal field at low temperature in case of Dy$_2$Ti$_2$O$_7$, but for Dy$_2$Ti$_{1.8}$Mn$_{0.2}$O$_7$ (FIG.\ref{Fig9}(b)), lattice volume decreases as per Debye equation till 70 K then it deviates and becomes somewhat constant then again it drops below 40 K this suggests that with Mn substitution crystal field prominence starts at higher temperature (70 K).
The obtained value of $V (T=0)$, $\theta_{D}$ and 9$\gamma$Nk$_{B}$/B , fitting parameters are depicted in the Table 1. It shows the enhancement in  Debye temperature with Mn inclusion at Ti site to $\theta_{D}$ = 290.6 for
Dy$_2$Ti$_{1.8}$Mn$_{0.2}$O$_7$ in comparison to $\theta_{D}$ = 214.1 for Dy$_2$Ti$_2$O$_7$. Such an enhancement shows the increase in the crystal field energy dominance to set in at higher temperature in case of Dy$_2$Ti$_{1.8}$Mn$_{0.2}$O$_7$ at 70 K.
\section{CONCLUSIONS}
We have studied low temperature magnetic ordering and spin dynamics for the polycrystalline compound Dy$_2$Ti$_{1.8}$Mn$_{0.2}$O$_7$. With small Mn inclusion at Ti site in Dy$_2$Ti$_2$O$_7$, We observe an emergence of a low-temperature spin ice freezing relaxation around 5 K (T$_i$) along with the second spin relaxation feature around 2.5 K (T \textless T$_i$) even in absence of any dc magnetic field which was not observed in case of Dy$_2$Ti$_2$O$_7$. Both the spin relaxations at T$_i$ and T \textless T$_i$ shows strong frequency dependence and Arrhenius fit shows that these spin relaxation are thermally driven. Mn is altering the spin dynamics of Dy$_2$Ti$_2$O$_7$ at low temperature regime. The low temperature Synchrotron x-ray diffraction study confirms absence of structural phase transition in this system. Debye parameters calculated by fitting Temperature-dependent variation of lattice volume using Debye Gruineisen equation. It shows that crystal field energy dominance set in at higher temperatures in case of Dy$_2$Ti$_{1.8}$Mn$_{0.2}$O$_7$ in comparison to Dy$_2$Ti$_2$O$_7$. Such a small perturbation in the composition makes this system suitable for applications, where spin ice like freezing needed to be explored at a workable temperatures.   
\section{ACKNOWLEDGEMENTS}
We acknowledge CIFC, IIT (BHU) for magnetic measurements. The Synchrotron XRD was carried out at PETRA III of DESY, a member of the Helmholtz Association (HGF). Financial support by the Department of Science and Technology (Government of India) provided within the framework of the India@DESY collaboration is gratefully acknowledged.

\begin{thebibliography}{28}%
\makeatletter
\providecommand \@ifxundefined [1]{%
 \@ifx{#1\undefined}
}%
\providecommand \@ifnum [1]{%
 \ifnum #1\expandafter \@firstoftwo
 \else \expandafter \@secondoftwo
 \fi
}%
\providecommand \@ifx [1]{%
 \ifx #1\expandafter \@firstoftwo
 \else \expandafter \@secondoftwo
 \fi
}%
\providecommand \natexlab [1]{#1}%
\providecommand \enquote  [1]{``#1''}%
\providecommand \bibnamefont  [1]{#1}%
\providecommand \bibfnamefont [1]{#1}%
\providecommand \citenamefont [1]{#1}%
\providecommand \href@noop [0]{\@secondoftwo}%
\providecommand \href [0]{\begingroup \@sanitize@url \@href}%
\providecommand \@href[1]{\@@startlink{#1}\@@href}%
\providecommand \@@href[1]{\endgroup#1\@@endlink}%
\providecommand \@sanitize@url [0]{\catcode `\\12\catcode `\$12\catcode
  `\&12\catcode `\#12\catcode `\^12\catcode `\_12\catcode `\%12\relax}%
\providecommand \@@startlink[1]{}%
\providecommand \@@endlink[0]{}%
\providecommand \url  [0]{\begingroup\@sanitize@url \@url }%
\providecommand \@url [1]{\endgroup\@href {#1}{\urlprefix }}%
\providecommand \urlprefix  [0]{URL }%
\providecommand \Eprint [0]{\href }%
\providecommand \doibase [0]{https://doi.org/}%
\providecommand \selectlanguage [0]{\@gobble}%
\providecommand \bibinfo  [0]{\@secondoftwo}%
\providecommand \bibfield  [0]{\@secondoftwo}%
\providecommand \translation [1]{[#1]}%
\providecommand \BibitemOpen [0]{}%
\providecommand \bibitemStop [0]{}%
\providecommand \bibitemNoStop [0]{.\EOS\space}%
\providecommand \EOS [0]{\spacefactor3000\relax}%
\providecommand \BibitemShut  [1]{\csname bibitem#1\endcsname}%
\let\auto@bib@innerbib\@empty
\bibitem [{\citenamefont {Subramanian}\ and\ \citenamefont
  {Sleight}(1993)}]{subramanian1993rare}%
  \BibitemOpen
  \bibfield  {author} {\bibinfo {author} {\bibfnamefont {M.}~\bibnamefont
  {Subramanian}}\ and\ \bibinfo {author} {\bibfnamefont {A.}~\bibnamefont
  {Sleight}},\ }\bibfield  {title} {\bibinfo {title} {Rare earth pyrochlores},\
  }\href@noop {} {\bibfield  {journal} {\bibinfo  {journal} {Handbook on the
  physics and chemistry of rare earths}\ }\textbf {\bibinfo {volume} {16}},\
  \bibinfo {pages} {225} (\bibinfo {year} {1993})}\BibitemShut {NoStop}%
\bibitem [{\citenamefont {Gardner}\ \emph {et~al.}(2010)\citenamefont
  {Gardner}, \citenamefont {Gingras},\ and\ \citenamefont
  {Greedan}}]{gardner2010magnetic}%
  \BibitemOpen
  \bibfield  {author} {\bibinfo {author} {\bibfnamefont {J.~S.}\ \bibnamefont
  {Gardner}}, \bibinfo {author} {\bibfnamefont {M.~J.}\ \bibnamefont
  {Gingras}},\ and\ \bibinfo {author} {\bibfnamefont {J.~E.}\ \bibnamefont
  {Greedan}},\ }\bibfield  {title} {\bibinfo {title} {Magnetic pyrochlore
  oxides},\ }\href@noop {} {\bibfield  {journal} {\bibinfo  {journal} {Rev.
  Mod. Phys.}\ }\textbf {\bibinfo {volume} {82}},\ \bibinfo {pages} {53}
  (\bibinfo {year} {2010})}\BibitemShut {NoStop}%
\bibitem [{\citenamefont {Greedan}(2006)}]{greedan2006frustrated}%
  \BibitemOpen
  \bibfield  {author} {\bibinfo {author} {\bibfnamefont {J.~E.}\ \bibnamefont
  {Greedan}},\ }\bibfield  {title} {\bibinfo {title} {Frustrated rare earth
  magnetism: Spin glasses, spin liquids and spin ices in pyrochlore oxides},\
  }\href@noop {} {\bibfield  {journal} {\bibinfo  {journal} {J. Alloys Compd.}\
  }\textbf {\bibinfo {volume} {408}},\ \bibinfo {pages} {444} (\bibinfo {year}
  {2006})}\BibitemShut {NoStop}%
\bibitem [{\citenamefont {Balents}(2010)}]{balents2010spin}%
  \BibitemOpen
  \bibfield  {author} {\bibinfo {author} {\bibfnamefont {L.}~\bibnamefont
  {Balents}},\ }\bibfield  {title} {\bibinfo {title} {Spin liquids in
  frustrated magnets},\ }\href@noop {} {\bibfield  {journal} {\bibinfo
  {journal} {Nature}\ }\textbf {\bibinfo {volume} {464}},\ \bibinfo {pages}
  {199} (\bibinfo {year} {2010})}\BibitemShut {NoStop}%
\bibitem [{\citenamefont {Harris}\ \emph {et~al.}(1997)\citenamefont {Harris},
  \citenamefont {Bramwell}, \citenamefont {McMorrow}, \citenamefont {Zeiske},\
  and\ \citenamefont {Godfrey}}]{harris1997geometrical}%
  \BibitemOpen
  \bibfield  {author} {\bibinfo {author} {\bibfnamefont {M.~J.}\ \bibnamefont
  {Harris}}, \bibinfo {author} {\bibfnamefont {S.}~\bibnamefont {Bramwell}},
  \bibinfo {author} {\bibfnamefont {D.}~\bibnamefont {McMorrow}}, \bibinfo
  {author} {\bibfnamefont {T.}~\bibnamefont {Zeiske}},\ and\ \bibinfo {author}
  {\bibfnamefont {K.}~\bibnamefont {Godfrey}},\ }\bibfield  {title} {\bibinfo
  {title} {Geometrical frustration in the ferromagnetic pyrochlore
  \ce{Ho$_2$Ti$_2$O$_7$}},\ }\href@noop {} {\bibfield  {journal} {\bibinfo
  {journal} {Phys. Rev. Lett.}\ }\textbf {\bibinfo {volume} {79}},\ \bibinfo
  {pages} {2554} (\bibinfo {year} {1997})}\BibitemShut {NoStop}%
\bibitem [{\citenamefont {Bramwell}\ and\ \citenamefont
  {Gingras}(2001)}]{bramwell2001spin}%
  \BibitemOpen
  \bibfield  {author} {\bibinfo {author} {\bibfnamefont {S.~T.}\ \bibnamefont
  {Bramwell}}\ and\ \bibinfo {author} {\bibfnamefont {M.~J.}\ \bibnamefont
  {Gingras}},\ }\bibfield  {title} {\bibinfo {title} {Spin ice state in
  frustrated magnetic pyrochlore materials},\ }\href@noop {} {\bibfield
  {journal} {\bibinfo  {journal} {Science}\ }\textbf {\bibinfo {volume}
  {294}},\ \bibinfo {pages} {1495} (\bibinfo {year} {2001})}\BibitemShut
  {NoStop}%
\bibitem [{\citenamefont {Bramwell}\ and\ \citenamefont
  {Harris}(2020)}]{bramwell2020history}%
  \BibitemOpen
  \bibfield  {author} {\bibinfo {author} {\bibfnamefont {S.~T.}\ \bibnamefont
  {Bramwell}}\ and\ \bibinfo {author} {\bibfnamefont {M.~J.}\ \bibnamefont
  {Harris}},\ }\bibfield  {title} {\bibinfo {title} {The history of spin ice},\
  }\href@noop {} {\bibfield  {journal} {\bibinfo  {journal} {J. Phys. Condens.
  Matter}\ }\textbf {\bibinfo {volume} {32}},\ \bibinfo {pages} {374010}
  (\bibinfo {year} {2020})}\BibitemShut {NoStop}%
\bibitem [{\citenamefont {Ramirez}\ \emph {et~al.}(1999)\citenamefont
  {Ramirez}, \citenamefont {Hayashi}, \citenamefont {Cava}, \citenamefont
  {Siddharthan},\ and\ \citenamefont {Shastry}}]{ramirez1999zero}%
  \BibitemOpen
  \bibfield  {author} {\bibinfo {author} {\bibfnamefont {A.~P.}\ \bibnamefont
  {Ramirez}}, \bibinfo {author} {\bibfnamefont {A.}~\bibnamefont {Hayashi}},
  \bibinfo {author} {\bibfnamefont {R.~J.}\ \bibnamefont {Cava}}, \bibinfo
  {author} {\bibfnamefont {R.}~\bibnamefont {Siddharthan}},\ and\ \bibinfo
  {author} {\bibfnamefont {B.}~\bibnamefont {Shastry}},\ }\bibfield  {title}
  {\bibinfo {title} {Zero-point entropy in ‘spin ice’},\ }\href@noop {}
  {\bibfield  {journal} {\bibinfo  {journal} {Nature}\ }\textbf {\bibinfo
  {volume} {399}},\ \bibinfo {pages} {333} (\bibinfo {year}
  {1999})}\BibitemShut {NoStop}%
\bibitem [{\citenamefont {Siddharthan}\ \emph {et~al.}(1999)\citenamefont
  {Siddharthan}, \citenamefont {Shastry}, \citenamefont {Ramirez},
  \citenamefont {Hayashi}, \citenamefont {Cava},\ and\ \citenamefont
  {Rosenkranz}}]{siddharthan1999ising}%
  \BibitemOpen
  \bibfield  {author} {\bibinfo {author} {\bibfnamefont {R.}~\bibnamefont
  {Siddharthan}}, \bibinfo {author} {\bibfnamefont {B.}~\bibnamefont
  {Shastry}}, \bibinfo {author} {\bibfnamefont {A.}~\bibnamefont {Ramirez}},
  \bibinfo {author} {\bibfnamefont {A.}~\bibnamefont {Hayashi}}, \bibinfo
  {author} {\bibfnamefont {R.}~\bibnamefont {Cava}},\ and\ \bibinfo {author}
  {\bibfnamefont {S.}~\bibnamefont {Rosenkranz}},\ }\bibfield  {title}
  {\bibinfo {title} {Ising pyrochlore magnets: Low-temperature
  properties,“ice rules,” and beyond},\ }\href@noop {} {\bibfield
  {journal} {\bibinfo  {journal} {Phys. Rev. Lett.}\ }\textbf {\bibinfo
  {volume} {83}},\ \bibinfo {pages} {1854} (\bibinfo {year}
  {1999})}\BibitemShut {NoStop}%
\bibitem [{\citenamefont {Matsuhira}\ \emph {et~al.}(2001)\citenamefont
  {Matsuhira}, \citenamefont {Hinatsu},\ and\ \citenamefont
  {Sakakibara}}]{matsuhira2001novel}%
  \BibitemOpen
  \bibfield  {author} {\bibinfo {author} {\bibfnamefont {K.}~\bibnamefont
  {Matsuhira}}, \bibinfo {author} {\bibfnamefont {Y.}~\bibnamefont {Hinatsu}},\
  and\ \bibinfo {author} {\bibfnamefont {T.}~\bibnamefont {Sakakibara}},\
  }\bibfield  {title} {\bibinfo {title} {Novel dynamical magnetic properties in
  the spin ice compound \ce{Dy$_2$Ti$_2$O$_7$}},\ }\href@noop {} {\bibfield
  {journal} {\bibinfo  {journal} {J. Phys. Condens. Matter}\ }\textbf {\bibinfo
  {volume} {13}},\ \bibinfo {pages} {L737} (\bibinfo {year}
  {2001})}\BibitemShut {NoStop}%
\bibitem [{\citenamefont {Rosenkranz}\ \emph {et~al.}(2000)\citenamefont
  {Rosenkranz}, \citenamefont {Ramirez}, \citenamefont {Hayashi}, \citenamefont
  {Cava}, \citenamefont {Siddharthan},\ and\ \citenamefont
  {Shastry}}]{rosenkranz2000crystal}%
  \BibitemOpen
  \bibfield  {author} {\bibinfo {author} {\bibfnamefont {S.}~\bibnamefont
  {Rosenkranz}}, \bibinfo {author} {\bibfnamefont {A.}~\bibnamefont {Ramirez}},
  \bibinfo {author} {\bibfnamefont {A.}~\bibnamefont {Hayashi}}, \bibinfo
  {author} {\bibfnamefont {R.}~\bibnamefont {Cava}}, \bibinfo {author}
  {\bibfnamefont {R.}~\bibnamefont {Siddharthan}},\ and\ \bibinfo {author}
  {\bibfnamefont {B.}~\bibnamefont {Shastry}},\ }\bibfield  {title} {\bibinfo
  {title} {Crystal-field interaction in the pyrochlore magnet
  \ce{Ho$_2$Ti$_2$O$_7$}},\ }\href@noop {} {\bibfield  {journal} {\bibinfo
  {journal} {J. Appl. Phys.}\ }\textbf {\bibinfo {volume} {87}},\ \bibinfo
  {pages} {5914} (\bibinfo {year} {2000})}\BibitemShut {NoStop}%
\bibitem [{\citenamefont {Snyder}\ \emph {et~al.}(2001)\citenamefont {Snyder},
  \citenamefont {Slusky}, \citenamefont {Cava},\ and\ \citenamefont
  {Schiffer}}]{snyder2001spin}%
  \BibitemOpen
  \bibfield  {author} {\bibinfo {author} {\bibfnamefont {J.}~\bibnamefont
  {Snyder}}, \bibinfo {author} {\bibfnamefont {J.}~\bibnamefont {Slusky}},
  \bibinfo {author} {\bibfnamefont {R.}~\bibnamefont {Cava}},\ and\ \bibinfo
  {author} {\bibfnamefont {P.}~\bibnamefont {Schiffer}},\ }\bibfield  {title}
  {\bibinfo {title} {How ‘spin ice’freezes},\ }\href@noop {} {\bibfield
  {journal} {\bibinfo  {journal} {Nature}\ }\textbf {\bibinfo {volume} {413}},\
  \bibinfo {pages} {48} (\bibinfo {year} {2001})}\BibitemShut {NoStop}%
\bibitem [{\citenamefont {Snyder}\ \emph
  {et~al.}(2004{\natexlab{a}})\citenamefont {Snyder}, \citenamefont {Ueland},
  \citenamefont {Slusky}, \citenamefont {Karunadasa}, \citenamefont {Cava},\
  and\ \citenamefont {Schiffer}}]{snyder2004low}%
  \BibitemOpen
  \bibfield  {author} {\bibinfo {author} {\bibfnamefont {J.}~\bibnamefont
  {Snyder}}, \bibinfo {author} {\bibfnamefont {B.}~\bibnamefont {Ueland}},
  \bibinfo {author} {\bibfnamefont {J.}~\bibnamefont {Slusky}}, \bibinfo
  {author} {\bibfnamefont {H.}~\bibnamefont {Karunadasa}}, \bibinfo {author}
  {\bibfnamefont {R.}~\bibnamefont {Cava}},\ and\ \bibinfo {author}
  {\bibfnamefont {P.}~\bibnamefont {Schiffer}},\ }\bibfield  {title} {\bibinfo
  {title} {Low-temperature spin freezing in the \ce{Dy$_2$Ti$_2$O$_7$} spin
  ice},\ }\href@noop {} {\bibfield  {journal} {\bibinfo  {journal} {Phys. Rev.
  B.}\ }\textbf {\bibinfo {volume} {69}},\ \bibinfo {pages} {064414} (\bibinfo
  {year} {2004}{\natexlab{a}})}\BibitemShut {NoStop}%
\bibitem [{\citenamefont {Snyder}\ \emph {et~al.}(2003)\citenamefont {Snyder},
  \citenamefont {Ueland}, \citenamefont {Slusky}, \citenamefont {Karunadasa},
  \citenamefont {Cava}, \citenamefont {Mizel},\ and\ \citenamefont
  {Schiffer}}]{snyder2003quantum}%
  \BibitemOpen
  \bibfield  {author} {\bibinfo {author} {\bibfnamefont {J.}~\bibnamefont
  {Snyder}}, \bibinfo {author} {\bibfnamefont {B.}~\bibnamefont {Ueland}},
  \bibinfo {author} {\bibfnamefont {J.}~\bibnamefont {Slusky}}, \bibinfo
  {author} {\bibfnamefont {H.}~\bibnamefont {Karunadasa}}, \bibinfo {author}
  {\bibfnamefont {R.~J.}\ \bibnamefont {Cava}}, \bibinfo {author}
  {\bibfnamefont {A.}~\bibnamefont {Mizel}},\ and\ \bibinfo {author}
  {\bibfnamefont {P.}~\bibnamefont {Schiffer}},\ }\bibfield  {title} {\bibinfo
  {title} {Quantum-classical reentrant relaxation crossover in
  \ce{Dy$_2$Ti$_2$O$_7$} spin ice},\ }\href@noop {} {\bibfield  {journal}
  {\bibinfo  {journal} {Phys. Rev. Lett.}\ }\textbf {\bibinfo {volume} {91}},\
  \bibinfo {pages} {107201} (\bibinfo {year} {2003})}\BibitemShut {NoStop}%
\bibitem [{\citenamefont {Snyder}\ \emph
  {et~al.}(2004{\natexlab{b}})\citenamefont {Snyder}, \citenamefont {Ueland},
  \citenamefont {Mizel}, \citenamefont {Slusky}, \citenamefont {Karunadasa},
  \citenamefont {Cava},\ and\ \citenamefont {Schiffer}}]{snyder2004quantum}%
  \BibitemOpen
  \bibfield  {author} {\bibinfo {author} {\bibfnamefont {J.}~\bibnamefont
  {Snyder}}, \bibinfo {author} {\bibfnamefont {B.}~\bibnamefont {Ueland}},
  \bibinfo {author} {\bibfnamefont {A.}~\bibnamefont {Mizel}}, \bibinfo
  {author} {\bibfnamefont {J.}~\bibnamefont {Slusky}}, \bibinfo {author}
  {\bibfnamefont {H.}~\bibnamefont {Karunadasa}}, \bibinfo {author}
  {\bibfnamefont {R.}~\bibnamefont {Cava}},\ and\ \bibinfo {author}
  {\bibfnamefont {P.}~\bibnamefont {Schiffer}},\ }\bibfield  {title} {\bibinfo
  {title} {Quantum and thermal spin relaxation in the diluted spin ice
  \ce{Dy$_{2- x}$M$_x$Ti$_2$O$_7$} (m= lu, y)},\ }\href@noop {} {\bibfield
  {journal} {\bibinfo  {journal} {Phys. Rev. B.}\ }\textbf {\bibinfo {volume}
  {70}},\ \bibinfo {pages} {184431} (\bibinfo {year}
  {2004}{\natexlab{b}})}\BibitemShut {NoStop}%
\bibitem [{\citenamefont {Shukla}\ \emph {et~al.}(2020)\citenamefont {Shukla},
  \citenamefont {Upadhyay}, \citenamefont {Tolkiehn},\ and\ \citenamefont
  {Upadhyay}}]{shukla2020robust}%
  \BibitemOpen
  \bibfield  {author} {\bibinfo {author} {\bibfnamefont {M.}~\bibnamefont
  {Shukla}}, \bibinfo {author} {\bibfnamefont {R.}~\bibnamefont {Upadhyay}},
  \bibinfo {author} {\bibfnamefont {M.}~\bibnamefont {Tolkiehn}},\ and\
  \bibinfo {author} {\bibfnamefont {C.}~\bibnamefont {Upadhyay}},\ }\bibfield
  {title} {\bibinfo {title} {Robust spin-ice freezing in magnetically
  frustrated \ce{Ho$_2$Ge$_x$Ti$_{2- x}$O$_7$} pyrochlore},\ }\href@noop {}
  {\bibfield  {journal} {\bibinfo  {journal} {J. Phys. Condens. Matter}\
  }\textbf {\bibinfo {volume} {32}},\ \bibinfo {pages} {465804} (\bibinfo
  {year} {2020})}\BibitemShut {NoStop}%
\bibitem [{\citenamefont
  {Rodr{\'\i}guez-Carvajal}(2001)}]{rodriguez2001introduction}%
  \BibitemOpen
  \bibfield  {author} {\bibinfo {author} {\bibfnamefont {J.}~\bibnamefont
  {Rodr{\'\i}guez-Carvajal}},\ }\bibfield  {title} {\bibinfo {title} {An
  introduction to the program fullprof},\ }\href@noop {} {\bibfield  {journal}
  {\bibinfo  {journal} {Laboratoire Leon Brillouin (CEA-CNRS)}\ } (\bibinfo
  {year} {2001})}\BibitemShut {NoStop}%
\bibitem [{\citenamefont {Liu}\ \emph {et~al.}(2014)\citenamefont {Liu},
  \citenamefont {Zou}, \citenamefont {Zhang}, \citenamefont {Ling},
  \citenamefont {Yu}, \citenamefont {He}, \citenamefont {Zhang},\ and\
  \citenamefont {Zhang}}]{liu2014magnetic}%
  \BibitemOpen
  \bibfield  {author} {\bibinfo {author} {\bibfnamefont {H.}~\bibnamefont
  {Liu}}, \bibinfo {author} {\bibfnamefont {Y.}~\bibnamefont {Zou}}, \bibinfo
  {author} {\bibfnamefont {L.}~\bibnamefont {Zhang}}, \bibinfo {author}
  {\bibfnamefont {L.}~\bibnamefont {Ling}}, \bibinfo {author} {\bibfnamefont
  {H.}~\bibnamefont {Yu}}, \bibinfo {author} {\bibfnamefont {L.}~\bibnamefont
  {He}}, \bibinfo {author} {\bibfnamefont {C.}~\bibnamefont {Zhang}},\ and\
  \bibinfo {author} {\bibfnamefont {Y.}~\bibnamefont {Zhang}},\ }\bibfield
  {title} {\bibinfo {title} {Magnetic order and dynamical properties of the
  spin-frustrated magnet \ce{Dy$_{2- x}$Yb$_x$Ti$_2$O$_7$}},\ }\href@noop {}
  {\bibfield  {journal} {\bibinfo  {journal} {J. Magn. Magn. Mater.}\ }\textbf
  {\bibinfo {volume} {349}},\ \bibinfo {pages} {173} (\bibinfo {year}
  {2014})}\BibitemShut {NoStop}%
\bibitem [{\citenamefont {Liu}\ \emph {et~al.}(2015)\citenamefont {Liu},
  \citenamefont {Zou}, \citenamefont {Ling}, \citenamefont {Zhang},
  \citenamefont {Zhang},\ and\ \citenamefont {Zhang}}]{liu2015enhanced}%
  \BibitemOpen
  \bibfield  {author} {\bibinfo {author} {\bibfnamefont {H.}~\bibnamefont
  {Liu}}, \bibinfo {author} {\bibfnamefont {Y.}~\bibnamefont {Zou}}, \bibinfo
  {author} {\bibfnamefont {L.}~\bibnamefont {Ling}}, \bibinfo {author}
  {\bibfnamefont {L.}~\bibnamefont {Zhang}}, \bibinfo {author} {\bibfnamefont
  {C.}~\bibnamefont {Zhang}},\ and\ \bibinfo {author} {\bibfnamefont
  {Y.}~\bibnamefont {Zhang}},\ }\bibfield  {title} {\bibinfo {title} {Enhanced
  ferromagnetism and emergence of spin-glass-like transition in pyrochlore
  compound \ce{Dy$_2$Ti$_{2- x}$V$_x$O$_7$}},\ }\href@noop {} {\bibfield
  {journal} {\bibinfo  {journal} {J. Magn. Magn. Mater.}\ }\textbf {\bibinfo
  {volume} {388}},\ \bibinfo {pages} {135} (\bibinfo {year}
  {2015})}\BibitemShut {NoStop}%
\bibitem [{\citenamefont {Anand}\ \emph {et~al.}(2015)\citenamefont {Anand},
  \citenamefont {Tennant},\ and\ \citenamefont
  {Lake}}]{anand2015investigations}%
  \BibitemOpen
  \bibfield  {author} {\bibinfo {author} {\bibfnamefont {V.}~\bibnamefont
  {Anand}}, \bibinfo {author} {\bibfnamefont {D.}~\bibnamefont {Tennant}},\
  and\ \bibinfo {author} {\bibfnamefont {B.}~\bibnamefont {Lake}},\ }\bibfield
  {title} {\bibinfo {title} {Investigations of the effect of nonmagnetic ca
  substitution for magnetic dy on spin-freezing in \ce{Dy$_2$Ti$_2$O$_7$}},\
  }\href@noop {} {\bibfield  {journal} {\bibinfo  {journal} {J. Phys. Condens.
  Matter}\ }\textbf {\bibinfo {volume} {27}},\ \bibinfo {pages} {436001}
  (\bibinfo {year} {2015})}\BibitemShut {NoStop}%
\bibitem [{\citenamefont {Fukazawa}\ \emph {et~al.}(2002)\citenamefont
  {Fukazawa}, \citenamefont {Melko}, \citenamefont {Higashinaka}, \citenamefont
  {Maeno},\ and\ \citenamefont {Gingras}}]{fukazawa2002magnetic}%
  \BibitemOpen
  \bibfield  {author} {\bibinfo {author} {\bibfnamefont {H.}~\bibnamefont
  {Fukazawa}}, \bibinfo {author} {\bibfnamefont {R.}~\bibnamefont {Melko}},
  \bibinfo {author} {\bibfnamefont {R.}~\bibnamefont {Higashinaka}}, \bibinfo
  {author} {\bibfnamefont {Y.}~\bibnamefont {Maeno}},\ and\ \bibinfo {author}
  {\bibfnamefont {M.}~\bibnamefont {Gingras}},\ }\bibfield  {title} {\bibinfo
  {title} {Magnetic anisotropy of the spin-ice compound
  \ce{Dy$_2$Ti$_2$O$_7$}},\ }\href@noop {} {\bibfield  {journal} {\bibinfo
  {journal} {Phys. Rev. B.}\ }\textbf {\bibinfo {volume} {65}},\ \bibinfo
  {pages} {054410} (\bibinfo {year} {2002})}\BibitemShut {NoStop}%
\bibitem [{\citenamefont {Ke}\ \emph {et~al.}(2007)\citenamefont {Ke},
  \citenamefont {Freitas}, \citenamefont {Ueland}, \citenamefont {Lau},
  \citenamefont {Dahlberg}, \citenamefont {Cava}, \citenamefont {Moessner},\
  and\ \citenamefont {Schiffer}}]{ke2007nonmonotonic}%
  \BibitemOpen
  \bibfield  {author} {\bibinfo {author} {\bibfnamefont {X.}~\bibnamefont
  {Ke}}, \bibinfo {author} {\bibfnamefont {R.}~\bibnamefont {Freitas}},
  \bibinfo {author} {\bibfnamefont {B.}~\bibnamefont {Ueland}}, \bibinfo
  {author} {\bibfnamefont {G.}~\bibnamefont {Lau}}, \bibinfo {author}
  {\bibfnamefont {M.}~\bibnamefont {Dahlberg}}, \bibinfo {author}
  {\bibfnamefont {R.}~\bibnamefont {Cava}}, \bibinfo {author} {\bibfnamefont
  {R.}~\bibnamefont {Moessner}},\ and\ \bibinfo {author} {\bibfnamefont
  {P.}~\bibnamefont {Schiffer}},\ }\bibfield  {title} {\bibinfo {title}
  {Nonmonotonic zero-point entropy in diluted spin ice},\ }\href@noop {}
  {\bibfield  {journal} {\bibinfo  {journal} {Phys. Rev. Lett.}\ }\textbf
  {\bibinfo {volume} {99}},\ \bibinfo {pages} {137203} (\bibinfo {year}
  {2007})}\BibitemShut {NoStop}%
\bibitem [{\citenamefont {Xing}\ \emph {et~al.}(2010)\citenamefont {Xing},
  \citenamefont {He}, \citenamefont {Feng}, \citenamefont {Guo}, \citenamefont
  {Zeng},\ and\ \citenamefont {Xu}}]{xing2010emergent}%
  \BibitemOpen
  \bibfield  {author} {\bibinfo {author} {\bibfnamefont {H.}~\bibnamefont
  {Xing}}, \bibinfo {author} {\bibfnamefont {M.}~\bibnamefont {He}}, \bibinfo
  {author} {\bibfnamefont {C.}~\bibnamefont {Feng}}, \bibinfo {author}
  {\bibfnamefont {H.}~\bibnamefont {Guo}}, \bibinfo {author} {\bibfnamefont
  {H.}~\bibnamefont {Zeng}},\ and\ \bibinfo {author} {\bibfnamefont {Z.-A.}\
  \bibnamefont {Xu}},\ }\bibfield  {title} {\bibinfo {title} {Emergent order in
  the spin-frustrated system \ce{Dy$_x$Tb$_{2- x}$Ti$_2$O$_7$} studied by ac
  susceptibility measurements},\ }\href@noop {} {\bibfield  {journal} {\bibinfo
   {journal} {Phys. Rev. B.}\ }\textbf {\bibinfo {volume} {81}},\ \bibinfo
  {pages} {134426} (\bibinfo {year} {2010})}\BibitemShut {NoStop}%
\bibitem [{\citenamefont {Gardner}\ \emph {et~al.}(1999)\citenamefont
  {Gardner}, \citenamefont {Gaulin}, \citenamefont {Lee}, \citenamefont
  {Broholm}, \citenamefont {Raju},\ and\ \citenamefont
  {Greedan}}]{gardner1999glassy}%
  \BibitemOpen
  \bibfield  {author} {\bibinfo {author} {\bibfnamefont {J.}~\bibnamefont
  {Gardner}}, \bibinfo {author} {\bibfnamefont {B.}~\bibnamefont {Gaulin}},
  \bibinfo {author} {\bibfnamefont {S.-H.}\ \bibnamefont {Lee}}, \bibinfo
  {author} {\bibfnamefont {C.}~\bibnamefont {Broholm}}, \bibinfo {author}
  {\bibfnamefont {N.}~\bibnamefont {Raju}},\ and\ \bibinfo {author}
  {\bibfnamefont {J.}~\bibnamefont {Greedan}},\ }\bibfield  {title} {\bibinfo
  {title} {Glassy statics and dynamics in the chemically ordered pyrochlore
  antiferromagnet \ce{Y$_2$Mo$_2$O$_7$}},\ }\href@noop {} {\bibfield  {journal}
  {\bibinfo  {journal} {Phys. Rev. Lett.}\ }\textbf {\bibinfo {volume} {83}},\
  \bibinfo {pages} {211} (\bibinfo {year} {1999})}\BibitemShut {NoStop}%
\bibitem [{\citenamefont {Mydosh}(1993)}]{mydosh1993spin}%
  \BibitemOpen
  \bibfield  {author} {\bibinfo {author} {\bibfnamefont {J.~A.}\ \bibnamefont
  {Mydosh}},\ }\href@noop {} {\emph {\bibinfo {title} {Spin glasses: an
  experimental introduction}}}\ (\bibinfo  {publisher} {CRC Press},\ \bibinfo
  {year} {1993})\BibitemShut {NoStop}%
\bibitem [{\citenamefont {Shi}\ \emph {et~al.}(2007)\citenamefont {Shi},
  \citenamefont {Tang}, \citenamefont {Zhu}, \citenamefont {Huang},
  \citenamefont {Yin}, \citenamefont {Li}, \citenamefont {Wang},\ and\
  \citenamefont {Wen}}]{shi2007dynamical}%
  \BibitemOpen
  \bibfield  {author} {\bibinfo {author} {\bibfnamefont {J.}~\bibnamefont
  {Shi}}, \bibinfo {author} {\bibfnamefont {Z.}~\bibnamefont {Tang}}, \bibinfo
  {author} {\bibfnamefont {B.}~\bibnamefont {Zhu}}, \bibinfo {author}
  {\bibfnamefont {P.}~\bibnamefont {Huang}}, \bibinfo {author} {\bibfnamefont
  {D.}~\bibnamefont {Yin}}, \bibinfo {author} {\bibfnamefont {C.}~\bibnamefont
  {Li}}, \bibinfo {author} {\bibfnamefont {Y.}~\bibnamefont {Wang}},\ and\
  \bibinfo {author} {\bibfnamefont {H.}~\bibnamefont {Wen}},\ }\bibfield
  {title} {\bibinfo {title} {Dynamical magnetic properties of the spin ice
  crystal \ce{Dy$_2$Ti$_2$O$_7$}},\ }\href@noop {} {\bibfield  {journal}
  {\bibinfo  {journal} {J. Magn. Magn. Mater.}\ }\textbf {\bibinfo {volume}
  {310}},\ \bibinfo {pages} {1322} (\bibinfo {year} {2007})}\BibitemShut
  {NoStop}%
\bibitem [{\citenamefont {Yakub}(2017)}]{yakub2017modified}%
  \BibitemOpen
  \bibfield  {author} {\bibinfo {author} {\bibfnamefont {E.}~\bibnamefont
  {Yakub}},\ }\bibfield  {title} {\bibinfo {title} {The modified
  debye--gr{\"u}neisen model for highly compressed diamond},\ }\href@noop {}
  {\bibfield  {journal} {\bibinfo  {journal} {J. Low Temp. Phys.}\ }\textbf
  {\bibinfo {volume} {187}},\ \bibinfo {pages} {20} (\bibinfo {year}
  {2017})}\BibitemShut {NoStop}%
\bibitem [{\citenamefont {Kiyama}\ \emph {et~al.}(1996)\citenamefont {Kiyama},
  \citenamefont {Yoshimura}, \citenamefont {Kosuge}, \citenamefont {Ikeda},\
  and\ \citenamefont {Bando}}]{kiyama1996invar}%
  \BibitemOpen
  \bibfield  {author} {\bibinfo {author} {\bibfnamefont {T.}~\bibnamefont
  {Kiyama}}, \bibinfo {author} {\bibfnamefont {K.}~\bibnamefont {Yoshimura}},
  \bibinfo {author} {\bibfnamefont {K.}~\bibnamefont {Kosuge}}, \bibinfo
  {author} {\bibfnamefont {Y.}~\bibnamefont {Ikeda}},\ and\ \bibinfo {author}
  {\bibfnamefont {Y.}~\bibnamefont {Bando}},\ }\bibfield  {title} {\bibinfo
  {title} {Invar effect of \ce{SrRuO$_3$}: Itinerant electron magnetism of
  \ce{Ru} 4d electrons},\ }\href@noop {} {\bibfield  {journal} {\bibinfo
  {journal} {Phys. Rev. B.}\ }\textbf {\bibinfo {volume} {54}},\ \bibinfo
  {pages} {R756} (\bibinfo {year} {1996})}\BibitemShut {NoStop}%
\end{thebibliography}
\end{document}